\documentstyle[twocolumn,epsf,aps]{revtex}
\draft
\begin{document}
\title{Stationary state in a two-temperature model with competing 
dynamics}
\author{Attila Szolnoki}
\address{Research Institute for Technical Physics and Materials Science, 
H-1525 Budapest,
POB 49, Hungary}
\address{
\centering{
\medskip \em
\begin{minipage}{15.4cm}
{}\qquad A two-dimensional half-filled lattice gas model with nearest-neighbor 
attractive interaction is studied where particles
are coupled to two thermal baths at different temperatures $T_1$
and $T_2$. The hopping of particles is governed by the heat bath at 
temperature $T_1$ with probability $p$ and the other heat bath 
$(T_2)$ with probability $1-p$ independently of the hopping
direction. On a square lattice the vertical and horizontal 
interfaces become unstable while interfaces are stable in the diagonal
directions. As a consequence, particles condense into a tilted 
square in the novel ordered state. The $p$-dependence of the
resulting nonequilibrium stationary state is studied 
by Monte Carlo simulation and dynamical mean-field
approximation as well.
%\pacs{\noindent PACS numbers: 02.70.Lq, 05.50.+q }
\end{minipage}
}}
\maketitle

\narrowtext

Nonequilibrium phase transitions have been studied extensively in the past
decade. One of the important questions to address is how nonequilibrium
constraints influence the order-disorder phase transition and the 
stationary
state. A widely studied example is the celebrated kinetic Ising model 
where
the nonequilibrium stationary states are produced by competing dynamics
\cite{RZ}. The competing dynamics can be combined Glauber (spin-flip)
processes at different temperatures \cite{Gar}, competition of the 
Glauber 
and the Kawasaki (spin-exchange) dynamics \cite{DeMasi}, or spin 
exchanges in different directions with different probabilities. The 
latter case 
({\it anisotropic} Kawasaki dynamics) may be interpreted as a driven
diffusive \cite{KLS}, or a two-temperature lattice gas model \cite{Maes}
depending on whether the spin exchanges in different directions are 
governed 
by an external field or two different temperatures. One can introduce the 
{\it isotropic} version of two-temperature lattice gas model in which
the hopping of particles (spin exchange) is governed by 
randomly applied heat baths at
different temperatures independently from hopping directions. 

Although it is one of the simplest model with competing dynamics, it has
not been studied yet. On can suspect that this kind of mixture of Kawasaki
dynamics does not result in relevant nonequilibrium behavior and the
system can be described by introducing the concept of an effective 
temperature. In fact, an earlier study of Ising model with competing
Glauber dynamics \cite{Gar} has concluded similar result.

In the present paper we study a two-dimensional lattice gas model
where particles are coupled to two thermal baths at different temperatures
independently of the hopping direction (isotropic Kawasaki dynamics). 
The Monte Carlo simulations demonstrate that the stationary state  
differs completely from those of the corresponding equilibrium model.
In spite of the mentioned expectations the model shows relevant
nonequilibrium behaviors.

We consider a two-dimensional lattice gas on a square lattice with 
$L \times L = N$
sites under periodic boundary conditions. The occupation variable $n_i$
at site $i$ takes the values $0$ (empty) or $1$ (occupied) and half-filled
occupation ($\sum_i n_i = N/2$) is assumed. The energy of the system is 
given by 
\begin{equation}
E = -J \sum_{(i,j)} n_i n_j \,\,,
\label{eq:energy}
\end{equation}
where the summation is over the nearest-neighbor pairs and $J>0$. The 
particles can jump to one of the empty nearest-neighbor sites with the 
hopping rate
\begin{equation}
W = p w(\Delta E,T_1) + (1-p) w(\Delta E,T_2)\,\,,
\label{eq:prob}
\end{equation}
where $\Delta E$ is the energy difference between the final and initial 
configurations. The probability $w(\Delta E,T_\alpha) = 
min [1,\exp(-\Delta E/T_\alpha)]$ 
is the familiar Metropolis rate ($\alpha = 1,2$) where the lattice 
constant
and the Boltzmann constant are chosen to be unity. The hopping rate 
defined
in Eq.~(\ref{eq:prob}) may be interpreted as a randomly chosen contact to
a thermal bath at temperature $T_1$ with probability $p$ and another
thermal bath at temperature $T_2$ with probability $1-p$. 

In the case of $T_1 = T_2$, evidently, the above defined model is 
equivalent 
to the standard kinetic Ising model which undergoes an order-disorder 
phase
transition at $T_c = 0.567$. In this half-filled system
the particles condense into a strip below $T_c$ to minimize the 
interfacial
energy. It should be noted that in the ordered state  
the interface can be oriented either horizontally or vertically, thus
this ordered phase violates the $x-y$ symmetry.
A suitable order parameter for characterizing the transition to 
strip-like 
order is the anisotropic squared magnetization \cite{KLS}
\begin{equation}
m = \sqrt{\vert \langle M^2_y \rangle - \langle M^2_x \rangle \vert} \,\,,
\label{eq:oop}
\end{equation}
where
\begin{equation}
M^2_y = {1 \over L} \sum_x [ {1 \over L} \sum_y (2n_{xy} - 1)]^2.
\end{equation}

Henceforth we will restrict ourselves to the case of $T_1 = 0$ and
$T_2 = \infty$. Obviously, for small value of $p$ the hopping of 
particles is mostly governed by the heat bath at temperature $T_2$, 
therefore the stationary state is expected to be disordered.
In the opposite case, the stationary state becomes ordered 
in the $p \rightarrow 1$ limit. To calculate the critical value $p_c$ of 
the 
order-disorder transition point, we employ the dynamical mean-field 
approach
suggested by Dickman \cite{Dic}. This method has been applied 
successfully 
in a number of other nonequilibrium models \cite{Santos,SSM,Grandi}. 
The value of $p_c$ can be obtained by the linear stability analysis of
the spatially homogeneous disordered phase.
In this approach the first step is to set up the master equations which 
describe the time evolution of probabilities of clusters, where the size 
of
clusters characterizes the level of approximation. Next, we determine the
stationary solution of the master equations by assuming disordered phase.
In the following a small density gradient is applied and 
the current generated in response to the density gradient is calculated. 
Decreasing the parameter $p$ the sign of the current changes is changed
at a given value which can be identified as the critical point. 
The results of these approximations are $p_c^{(2p)} = 0.893$ at the 
two-point and $p_c^{(4p)} = 0.907$ at the four-point levels.

Monte Carlo simulation has been carried out to check the validity of the 
above predictions. 
We have used independent random
numbers for choosing what heat bath to couple the particle
to and for comparing with the corresponding hopping 
probability during an elementary Monte Carlo step. 
However, the qualitative behavior remained unchanged if the same random 
number was used for the above mentioned two steps.
The simulation were started from a perfectly ordered strip in
the presumed ordered region (at $p = 0.97$). During the simulation we 
have
monitored the  relaxation of order parameter defined in 
Eq.~(\ref{eq:oop}).  
Comparing the results of different system sizes, a puzzling behavior is
observed.
Namely, the stationary value of the order parameter 
decreases and tends to zero if we increase the system
size. To clarify this feature we have written a computer program
displaying the time evolution of configuration.

\begin{figure}
%\vspace{4cm}
\centerline{\epsfxsize=8.0cm
                   \epsfbox{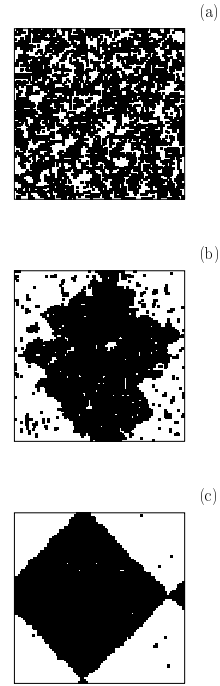}
                   \vspace*{2mm}      }
\caption{Typical configurations for a $100 \times 100$ system, at $p = 
0.85$ (a); $p = 0.95$ (b); and $p = 0.99$ (c).}
\label{fig:conf}
\end{figure}

This visualization of the particle configurations 
has indicated that the nonequilibrium condition influences the stability 
of interfaces. Namely, the interface 
in the (01) and (10) directions became unstable. At the same time,
the interfaces in the (11) and 3 other symmetrically equivalent 
directions proved to be stable. Consequently, the particles condense into
a tilted square in contrary to the strips observed in equilibrium system.
In Fig.~\ref{fig:conf} some typical configurations are
shown at different values of the control parameter $p$.
The titled square is the real stationary state because
the system evolves into this state from either vertically or horizontally
oriented strip. The opposite evolution has never been observed.
However, the necessary time ($\tau_E$) to evolve from "strip 
configuration"
to the "tilted square" may be rather long. As an example,
$\tau_E \approx 4 \times 10^6$ Monte Carlo steps for a system size
$L \times L = 100 \times 100$ and at $p = 0.97$. 

We have concluded that the order parameter defined by Eq.~(\ref{eq:oop})
cannot describe the novel type of the ordering process which becomes 
striking 
especially for large systems. An adequate order parameter for the
new nonequilibrium state can be defined as 

\begin{figure}
%\vspace{4cm}
\centerline{\epsfxsize=8.0cm
                   \epsfbox{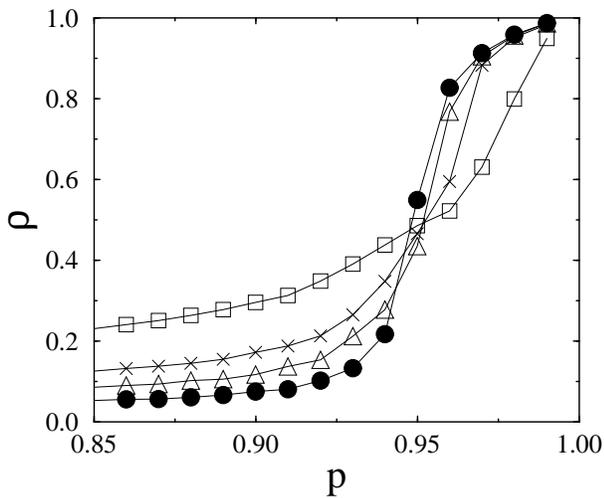}
                   \vspace*{2mm}      }
\caption{The 'new' order parameter as a function of $p$ for different
system
sizes. System sizes are $20 \times 20$ ($\Box$), $40 \times 40$
($\times$),
 $60 \times 60$ ($\triangle$), and $100 \times 100$ ($\bullet$).}
\label{fig:orderp}
\end{figure}

\begin{equation}
\rho = m_x \times m_y \,\,,
\label{eq:uop}
\end{equation}
where
\begin{equation}
m_x = {4 \over L^2} \sum_y \vert\sum_x (n
_{xy} - {1 \over
2})\vert.
\end{equation}
Using this definition we can describe the ordering process 
as demonstrated
in Fig.~\ref{fig:orderp}. As the new ordered state differs 
from the
corresponding equilibrium ordered phase, our system cannot
be described by the equilibrium model with an effective 
temperature. The explanation of instability of horizontal
(vertical) interface is related to the material transport
along the domain interface. To understand the
microscopic mechanism for this effect, it is instructive
to compare a horizontal and a diagonal oriented interface.
Suppose, a particle jumps out from a horizontal interface
in consequence of fluctuations and leaves a hole in the 
initial site. This particle can easily move along the
horizontal interface since there is no energy difference
between an initial and a final site. 
If the system size is large enough, the particle (hole)
may meet another particle (hole) and it initiates the
break-up of the interface. A significant difference has 
been detected in the movement of particles 
along the diagonal oriented interface. 
Here, jumps are blocked and the material transport is
reduced leaving the interface unchanged.
It is an interesting question
how a modification of the dynamics influences the stability 
of the diagonal interface. The movement along the interface
can be reduced to only one jump by allowing for 
next-nearest-neighbor jump as well. Now, the move along
the interface occurs with probability one (since 
$\Delta E = 0$),
similarly to the case of horizontal interface. As a
consequence, the diagonal orientation is not selected by
the interfacial mobility and the horizontal (or vertical)
direction, which contains lower interface energy, may be
preferred. To test this argument, we have performed MC
simulation on the modified model and the equilibrium 
strip-like state is found to be stable. We should mention
that diagonal oriented interface, which ensures minimum
excess interfacial energy on a square lattice has been obtained 
in a phase separation in chemically reactive mixtures \cite{Glotzer}.
A new type of
stationary state, as a consequence of nonequilibrium conditions, has
already been observed in other systems. For example, in a ferromagnetic 
Ising
system with competing Glauber and Kawasaki dynamics the stationary state
is identified with the antiferromagnetic state in a special parameter 
regime \cite{Tom}.

Returning to our model, we can define the derivative of 
energy with respect to the control parameter
$p$ similarly to the specific heat for equilibrium models. The quantity 
$C_p = \partial E / \partial p$ behaves like the equilibrium specific 
heat. 
The location of the maximum in $C_p$ can be identified as a transition 
point for a finite lattice.
Plotting the location of $C_p$ peak against $L^{-1}$, the linear fit
yields $p_c = 0.947(5)$ in the thermodynamic limit. This numerical
result agrees very well with the prediction of dynamical mean-field
approximation at four-point level (the difference is only $4 \%$).

Finally, we turn now to the problem of critical behavior briefly. A 
possible 
method to determine the critical indexes of a continuous phase 
transition is the finite-size scaling which has often yielded useful 
result
for nonequilibrium models \cite{Xu,Mar,Szol}. 
In the following we assume that the order parameter depends on the 
system size and the distance from the critical point as
\begin{equation}
\rho \sim L^{-\beta/\nu} f((p-p_c) L^{1/\nu})\,\,,
\label{eq:fss}
\end{equation}
where $\beta$ and $\nu$ are the exponents of the order parameter and
correlation-length.
Monte Carlo data for the order parameter are fitted to the scaling
form (\ref{eq:fss}) with the Ising exponents and we have found good data
collapse. This result is in agreement with the conjecture of Grinstein
{\it et. al.} for nonequilibrium ferromagnetic spin models
with up-down symmetry \cite{Grin}.

In summary, we have shown that the isotropic combination 
of the Kawasaki dynamics for two temperatures on a square
lattice can result in nontrivial behaviors in
nonequilibrium stationary state. At a critical value of the control 
parameter
$p_c$ the system segregates into a high-density "liquid" and a low-density
"gas" phase. However, in the stationary state the energy 
of the interface is higher than those of the corresponding equilibrium 
model. In the stationary state the diagonal interfaces become 
preferred to the horizontal and vertical ones. 
The phase transition describable by using a new suitable 
order parameter belongs to the Ising universality class. 
The stability of interfaces are related to the mobility
of particles along the interfaces, where the diagonal
orientation minimizes the influence of the energy flow
between the two heat baths. 
Although the stability of interfaces may be tied to the type
of lattice, the study of the corresponding
coarse-grained macroscopic model would be useful. 
However, there is no
straightforward way to find the macroscopic counterpart
of a microscopic model. There are examples where the 
microscopic and supposed macroscopic model yield different
morphology \cite{Alex}.
Nevertheless, we believe that the behavior of our model
is part of the general phenomenon where
the external energy input 
results in interfacial effects modifying the morphology of the
resulting stationary state \cite{Sab,polyd}. Further
work is required to clarify the connection between the suggested
model and the above mentioned driven nonequilibrium models.

\vspace{1cm}
The author thanks Gy\"orgy Szab\'o for his critical reading of the 
manuscript
and Ole G. Mouritsen who inspired this study. This research was 
supported by the Hungarian National Research Fund (OTKA) 
under Grant Nos. F-19560 and F-30449.

\end{document}